\documentclass[a4paper,11pt]{article}
\usepackage{amsfonts,amsmath,amssymb}
\usepackage{bm,tensor}
\usepackage{graphicx}
\usepackage{braket}
\usepackage{ifpdf}
\usepackage[utf8]{inputenc}
\usepackage{color}
\usepackage{physics}
\usepackage{url}
\usepackage{jcappub}
\usepackage{mathtools}
\usepackage{subfigure}

\usepackage[colorlinks=true
,urlcolor=blue
,anchorcolor=blue
,citecolor=blue
,filecolor=blue
,linkcolor=blue
,menucolor=blue
,linktocpage=true
,pdfproducer=medialab
,pdfa=true
]{hyperref}

\definecolor{dgreen}{rgb}{0.05, 0.60, 0.05}

\usepackage{ulem}

\title{Anisotropies in Cosmological 21~cm Background by Oscillons/I-balls of Ultra-light Axion-like Particle}
\author[a,b]{Masahiro Kawasaki,}
\author[a]{Kazuyoshi Miyazaki,}
\author[a]{Kai Murai,}
\author[a]{Hiromasa Nakatsuka,}
\author[a]{and Eisuke Sonomoto}

\affiliation[a]{ICRR, The University of Tokyo, Kashiwa, 277-8582, Japan}
\affiliation[b]{Kavli IPMU(WPI), UTIAS, University of Tokyo,Kashiwa,277-8583, Japan}
\emailAdd{kawasaki@icrr.u-tokyo.ac.jp}
\emailAdd{kmiyazak@icrr.u-tokyo.ac.jp}
\emailAdd{kmurai@icrr.u-tokyo.ac.jp}
\emailAdd{hiromasa@icrr.u-tokyo.ac.jp}

\abstract{
Ultra-light axion-like particle (ULAP) with mass $m \sim 10^{-22} ~\mathrm{eV}$ has recently been attracting attention as a possible solution to the small-scale crisis.
ULAP forms quasi-stable objects called oscillons/I-balls, which can survive up to a redshift $z \sim 10$ and affect the structure formation on a scale $\sim \mathcal{O}(0.1)~\mathrm{Mpc}$ by amplifying the density fluctuations.
We study the effect of oscillons on 21~cm anisotropies caused by neutral hydrogen in minihalos. 
It is found that this effect can be observed in a wide mass range by future observations such as Square Kilometer Array~(SKA) 
if the fraction of ULAP to the total dark matter density is $\mathcal{O}(0.01 \text{--} 0.1)$.
}
\keywords{Axion-like particle, 21~cm line, Oscillon}
\arxivnumber{}
\begin{document}
\maketitle



\section{Introduction}

Dark matter (DM) accounts for about 25\% of energy in the current universe, while its origin remains unexplained~\cite{1991MNRAS.249..523B,36.1.435,Koopmans:2002qh}.
The $\Lambda$CDM model
successfully explains observations of the cosmic microwave background fluctuations and the large-scale structures of the universe~\cite{Hinshaw:2013,DES:2017myr,Planck:2018vyg}.
However, when focusing on small scales, the numerical simulations based on the $\Lambda$CDM model show results contrary to some observations, giving rise to three cosmological problems; missing satellite problem~\cite{Klypin:1999uc,Moore:1999nt}, core-cusp problem~\cite{Navarro:1996gj,Moore:1999gc}, and too-big-to-fail problem~\cite{2012MNRAS.422.1203B,Bullock:2017xww}.
These problems
could be solved by ultralight bosonic DM with a mass of $\sim10^{-22}$~eV~\cite{Hu:2000ke,Hui:2016ltb}, which suppresses density fluctuations on small scales and behaves like CDM on large scales. 

Ultra-Light Axion-like Particle~(ULAP), which is predicted from the string theory~\cite{Svrcek:2006yi}, has been attracting attention as a possible solution to the small-scale problems.
In the early universe, the ULAP field starts oscillation and the oscillating ULAP field has spatial instabilities which generate local objects called oscillons/I-balls if the potential is shallower than quadratic~\cite{1976ZhPmR..24...15B,Gleiser:1993pt,Copeland:1995fq}.
Oscillons are quasi-stable due to their approximate conservation of the adiabatic invariant, and they could survive until the present time depending on the ULAP mass and the shape of the potential~\cite{Kasuya:2002zs,Kawasaki:2015vga,Ibe:2019lzv}.
Since oscillons are randomly produced, they generate the Poisson fluctuations on the ULAP density power spectrum.
Thus, if oscillons exist, they can affect the structure of the universe at scale $\sim \mathcal{O}(0.1)$~Mpc. 

The 21~cm line, which is produced by the hyperfine structure of neutral hydrogen, is useful to probe density fluctuation on small scales. 
Some works investigated the 21~cm line to constrain the axions~\cite{Kadota:2020ybe,Shimabukuro:2019gzu,Shimabukuro:2020tbs}, warm dark matter~\cite{Sekiguchi:2014wfa}, primordial black holes~\cite{Gong:2018sos,Villanueva-Domingo:2021cgh}, and oscillons~\cite{Kawasaki:2020tbo}.
Ref.~\cite{Kawasaki:2020tbo} discussed the halo formation including the effects of oscillons and evaluated the isotropic 21~cm line signal.
However, the anisotropies on the 21~cm line have not yet been investigated.

In this paper, we study anisotropies in the cosmological 21~cm background induced by oscillons made of the ULAP field. 
We assume that the ULAP accounts for a fraction of the total DM. 
We consider the coexistence of the oscillons and homogeneous ULAP following the numerical simulation~\cite{Kawasaki:2020jnw}.
We found that future experiments such as Square Kilometer Array (SKA)~\cite{SKA} 
can detect the effect of oscillons on the 21~cm line.


The structure of this paper is as follows. 
In Sec.~\ref{sec2}, we consider the matter power spectrum when ULAP partially accounts for the DM. 
We analytically calculate the matter power spectrum in the presence of oscillons 
and confirm that the oscillons enhance the fluctuations on small scales.
In Sec.~\ref{sec3}, we derive the 21~cm fluctuation. 
We will see that oscillons
affect the 21 cm fluctuations 
by changing the number density and formation time of minihalos.
In Sec.~\ref{sec4}, we discuss the detectability of the 21~cm fluctuation with oscillons by future observations.
We conclude the results in Sec.~\ref{sec5}.
All cosmological parameters in this paper are extracted from the results of Planck 2018~\cite{Planck:2018vyg}. 

\section{Matter Power Spectrum of ULAP}
\label{sec2}

In this paper, we consider that the dark matter consists of two components, unknown cold dark matter and ULAP.
The ULAP is further decomposed into two parts, (almost) homogeneous ULAP and ULAP oscillons.\footnote{
We use the term ``homogeneous'' to describe the ULAP component that does not form oscillons, including perturbations around the homogeneous component. 
} 
We define the density parameter of each component as 
\begin{align}
    \Omega_{\mathrm{DM}} 
     &= \Omega_{\mathrm{CDM}} + \Omega_{\mathrm{ULAP}} \\
     &= \Omega_{\mathrm{CDM}} + (\Omega_{\mathrm{homo}} + \Omega_{\mathrm{osc}}),
\end{align}
where $\Omega_\mathrm{CDM}$ is the density parameter of the CDM besides ULAP, and $\Omega_{\mathrm{ULAP}}$ represents that of the total ULAP including the homogeneous part $\Omega_{\mathrm{homo}}$ and oscillon part $\Omega_{\mathrm{osc}}$.
We also define the ratio of ULAP to DM and the ratio of oscillons to the total
ULAP as
\begin{align}
    f_{\mathrm{ULAP}}
    &\equiv 
    \frac{\Omega_{\mathrm{ULAP}}}{\Omega_{\mathrm{DM}}}
    \quad,\quad
     r_{\mathrm{osc}} 
    \equiv 
    \frac{\Omega_{\mathrm{osc}}}{\Omega_{\mathrm{ULAP}}}.
\end{align}

The oscillon energy fraction depends on the ULAP potential.
Here, we adopt the monodromy-type potential~\cite{Silverstein:2008sg,McAllister:2008hb,Nomura:2017ehb}, which is written as
\begin{equation}
    V(\phi)
    =
    \frac{m_a^{2} F^{2}}{2 p}
    \left[
        1 - \left( 1+\frac{\phi^{2}}{F^{2}} \right)^{-p}
    \right],
\end{equation}
where $m_a$ is the ULAP mass, and $F$ is the ULAP decay constant.
Oscillons are formed only when the potential is shallower than the quadratic term, $p>-1$~\cite{Kawasaki:2019czd}.
Due to the approximate conservation of adiabatic invariant, they can exist for a long time and can survive long enough until the epoch of our interest ($z\sim10$)~\cite{Ibe:2019vyo,Zhang:2020bec,Olle:2019kbo}.
Since the oscillon mass is related to the ULAP mass by $M_\mathrm{osc}\propto F^2/m_a$, a tiny ULAP mass leads to heavy oscillons.
For example, when $m\sim 10^{-22}$~eV and $F\sim 10^{15} $~GeV, the oscillon mass is $M_\text{osc}\sim 10^6\, M_\odot$.
In such a case, the number density of the oscillons is so small that their Poisson fluctuations cannot be neglected.
For this reason, the oscillons can affect the matter power spectrum.
The formed oscillons contain at most about 70\% of the total energy of the ULAP~\cite{Kawasaki:2020jnw}, and the rest exists as the homogeneous component of the ULAP.   

The homogeneous ULAP and oscillons affect the matter power spectrum differently.
The homogeneous ULAP suppresses the fluctuations on scales smaller than its de Broglie length.
On the other hand, oscillons induce the Poisson fluctuations.
In the following section, we quantitatively estimate the matter power spectrum taking these effects into account.

\subsection{Matter Power Spectrum without Oscillons}

When ULAP does not form oscillons or oscillons totally decay, we can decompose
the total matter power spectrum in the following form:
\begin{equation}
    P(k,t) = P_{\mathrm{CDM}}(k,t)+P_{\mathrm{ULAP}}(k,t),
    \label{eq: CDM and ULAP power spectrum}
\end{equation}
where $P_{\mathrm{ULAP}}(k,t)$ denotes the matter power spectrum of ULAP.
We calculate the power spectrum without oscillons using the \texttt{AxionCAMB} code~\cite{Hlozek:2014lca}. 
The calculation results are shown in Fig.~\ref{power_spectrum} as the dotted lines.
Since the ULAP without oscillons suppresses the small-scale fluctuations, the matter power spectrum is also suppressed compared with the $\Lambda$CDM power spectrum at $k \gtrsim \mathcal{O}(10)~\mathrm{Mpc}^{-1}$,
which can be seen in Fig.~\ref{power_spectrum}. 

\subsection{Matter Power Spectrum with Oscillons}
\label{sec:oscillon_power}

Next, let us show the oscillon matter power spectrum $P_{\mathrm{osc}}(k,t)$ following Ref.~\cite{Kawasaki:2020jnw}.
Assuming that the positions of oscillons are not correlated and neglecting the oscillon size ($k/a \ll m_a$),
the matter power spectrum of oscillons in the matter dominated era 
is written as
\begin{equation}
    P_{\mathrm{osc}}(k, t)
    =
    \left( 
        \frac{3a}{2a_{\mathrm{eq}}}
    \right)^2
    \frac{\left(r_{\mathrm{osc}} f_{\mathrm{ULAP}} \right)^{2}}{n_{\mathrm{osc}} a^{3}}
    \left(
        \frac{\Omega_{\mathrm{DM}}}{\Omega_{m}}
    \right)^{2}
    \frac{\left\langle M_{\mathrm{osc}}^{2} \right\rangle}{\left\langle M_{\mathrm{osc}}\right\rangle^{2}}
    \left[
        1-\left( \frac{2}{k L_{s}} \right)^{2} \sin^{2} \frac{k L_{s}}{2}
    \right],
\end{equation}
where $a_{\mathrm{eq}}$ is the scale factor at the matter-radiation equality, 
the bracket $\langle ~\rangle$ represents the ensemble average over the mass distribution of oscillons, $M_{\mathrm{osc}}$ is the mass of an oscillon, $n_{\mathrm{osc}}$ is the physical number density of oscillons, and $L_s$ is the cut-off scale due to energy conservation, which is determined by a simulation.
Because of the gravitational growth of the isocurvature fluctuations and the evaporation of oscillons, the oscillon power spectrum evolves over time.
The prefactor proportional to $a^2$ reflects the linear growth of the fluctuations after the matter-radiation equality.

In the presence of oscillons, the power spectrum is contributed by the Poisson fluctuations of oscillons in addition to that in Eq.~\eqref{eq: CDM and ULAP power spectrum}.
When evaluating the contribution of Eq.~\eqref{eq: CDM and ULAP power spectrum} by \texttt{AxionCAMB}, we set the homogeneous ULAP fraction as $f_{\mathrm{ULAP}}(1-r_{\mathrm{osc}})$ and consider oscillons as CDM.\footnote{
Although the oscillon fraction $r_{\mathrm{osc}}$ gradually decreases, $r_{\mathrm{osc}}$ does not change that much during the time we are interested in.
Considering this point, we used $r_{\mathrm{osc}}$ averaged over $z=10$ to $20$ in  \texttt{AxionCAMB}.
}
On the other hand, we evaluate the power spectrum of the Poisson fluctuations of oscillons following Refs.~\cite{Kawasaki:2020jnw,Kawasaki:2020tbo}.
Since the oscillon fraction $r_\mathrm{osc}$, number density $n_{\mathrm{osc}}$, and mean mass $\langle M_\mathrm{osc}\rangle$ gradually change due to the evaporation of oscillons, we follow the time evolution of oscillons to obtain such parameters for a fixed $z$.
The initial mass distribution of oscillons is determined by the numerical lattice simulation of the oscillon formation.
In this paper, we use the result of the lattice simulation in Ref.~\cite{Kawasaki:2020tbo}. (See appendix~B of Ref.~\cite{Kawasaki:2020tbo} for the setup of the numerical calculation.) 
After the formation, the subsequent time evolution of the mass distribution is calculated using the analytical decay rate of an oscillon~\cite{Zhang:2020bec,Ibe:2019vyo},
\begin{equation}
\label{gamma}
    \Gamma
    \equiv
    \frac{1}{M_{\mathrm{osc}}}
    \left|
        \frac{\mathrm{d} M_{\mathrm{osc}}}{\mathrm{d} t}
    \right|.
\end{equation}
The RHS of Eq.~\eqref{gamma} is evaluated by decomposing the ULAP field into the oscillon profile and perturbation part
$\phi(t, \bm{x}) = \phi_{\mathrm{osc}}(t, \bm{x}) + \xi(t, \bm{x})$,
and analytically solving the equation of motion of the perturbation part.
Then, the energy emission from an oscillon is given by the time average of the Poynting vector of the perturbation part
\begin{equation}
    \frac{\mathrm{d}M_{\mathrm{osc}}}{\mathrm{d}t}
    =
    4 \pi r^2 \overline{\partial_t \xi(t,r) \partial_t \xi(t,r)},
\end{equation}
where we assume that $\xi$ is spherically symmetric and the overline represents a time average.

\begin{figure}[tb]
    \centering
    \includegraphics[clip,width=15cm]{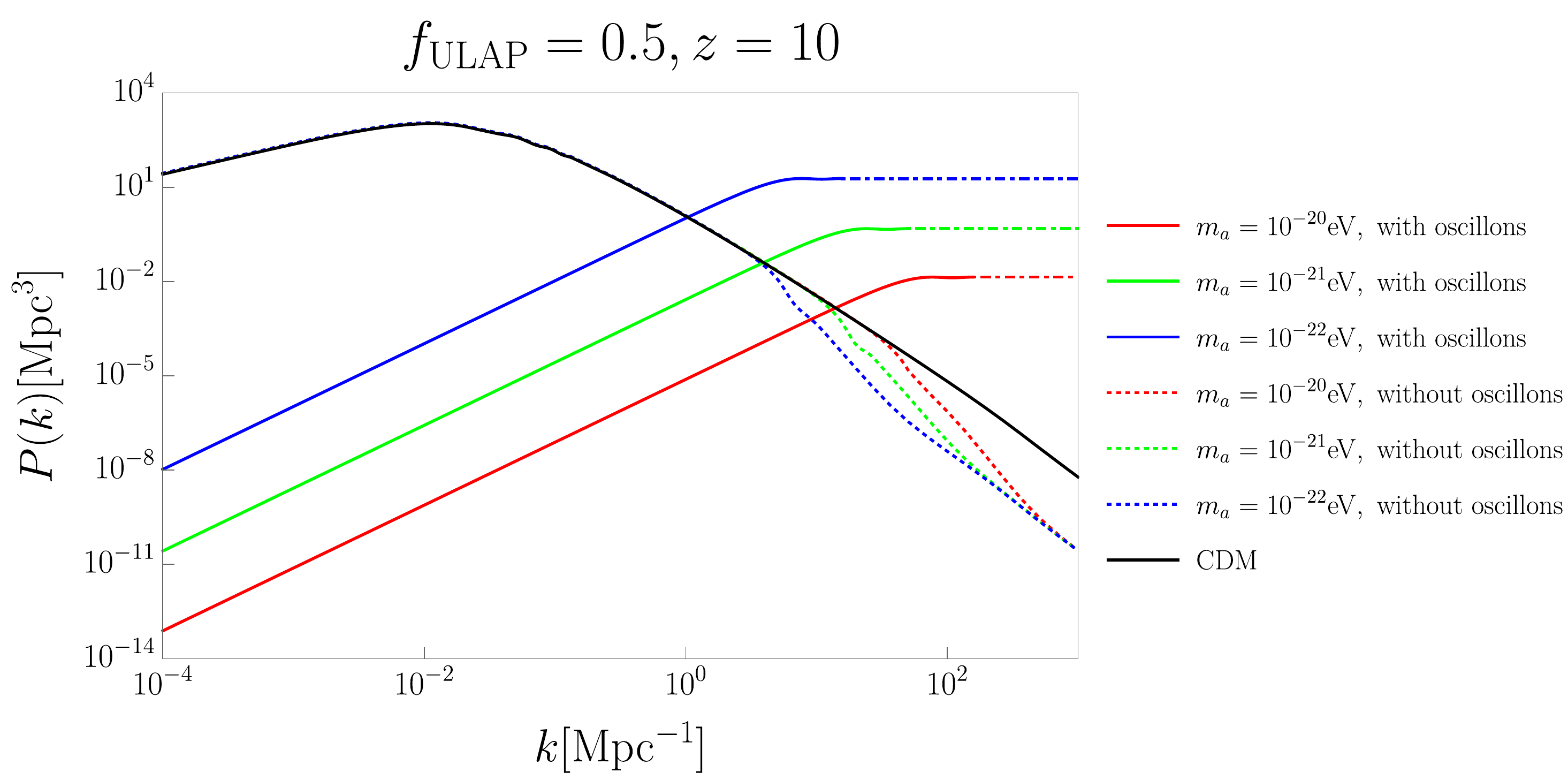}
    \caption{Matter power spectrum at $z=10$ for $f_{\mathrm{ULAP}}=0.5$. 
    The blue, green, and red solid lines show the oscillon matter power spectra for $m_a=10^{-20}~\mathrm{eV}, \, 10^{-21}~\mathrm{eV},$ and $10^{-22}~\mathrm{eV}$, respectively. 
    The dotted lines are the matter power spectra for the case of homogeneous ULAP, and the color difference is the same as in the previous case.
    We cut off the oscillon matter power spectra on the scale where the number of oscillons equals 10 because the fluctuations are non-linear.
    Corresponding scales are plotted with the dot-dashed lines.
    The matter power spectrum in the $\Lambda$CDM model is plotted with the black solid line for comparison.}
    \label{power_spectrum}
\end{figure}

The oscillon matter power spectrum for the ULAP mass $m_a=10^{-22}~\mathrm{eV}, \,10^{-21}~\mathrm{eV},$ and $10^{-20}~\mathrm{eV}$ are plotted in Fig.~\ref{power_spectrum} by the solid lines.
In the presence of oscillons, the small scale fluctuations are amplified and so is the matter power spectrum as can be seen in Fig~\ref{power_spectrum}.
We should note that, when one considers a single oscillon, the expression of $P_{\mathrm{osc}}$ is invalid because the fluctuations are non-linear on the scale containing a single oscillon.
So, we cut off the oscillon matter power spectrum on the scale $k_{\mathrm{c}}$ where the number of oscillons equals to 10, or equivalently $n_{\mathrm{osc}}(2\pi / k_{c})^3=10$.
Corresponding regions with $k > k_\mathrm{c}$ are plotted with the dot-dashed lines.
We have confirmed that changing this criterion from 10 to 2 does not largely change the final result.

\section{Minihalo and 21~cm Fluctuations}
\label{sec3}

In this section, following Ref.~\cite{Furlanetto:2002ng,Shimabukuro:2019gzu,Shimabukuro:2020tbs,Shimabukuro:2014ava}, we formulate
the properties of minihalos and the 21~cm fluctuations from minihalos in the presence of ULAP, which are different from the $\Lambda$CDM case. 

\subsection{Minihalo Mass Function}
\label{massfunction} 

\begin{figure}[t]
    \centering
    \includegraphics[width=15cm]{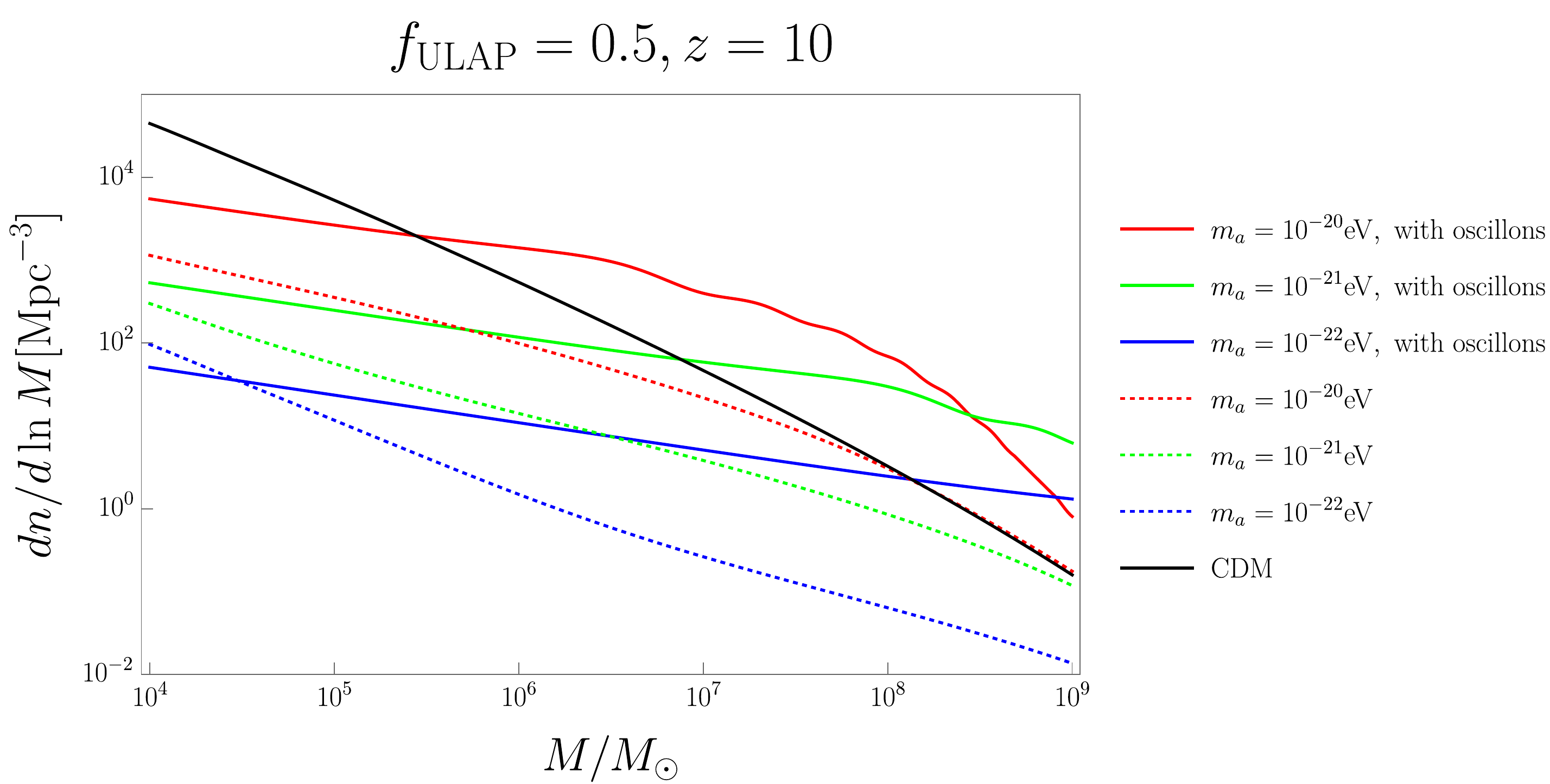}
    \caption{Minihalo mass functions for $m=10^{-20}$~eV (red),  $m=10^{-21}$~eV (Green), $m=10^{-22}$~eV (Blue) are shown.
    The solid and dotted lines denote the halo mass functions with and without  oscillons, respectively.
    We also show the case of the $\Lambda$CDM model with the black solid line for comparison.}
    \label{fig:massfunction}
\end{figure}

First, we introduce the comoving number density of minihalos per mass interval $M \sim M+\mathrm{d}M$, denoted by $\mathrm{d}n/\mathrm{d}M$.
In calculating the halo mass function, the Press-Schechter formalism is often used in the literature. 
In this paper, we adopt a more accurate fitting formula, Sheth-Tormen mass function~\cite{Sheth:1999mn}\footnote{
Exactly speaking, the Sheth-Tormen mass function cannot be used for the Poisson fluctuations of the oscillons because it assumes the Gaussian distribution for density fluctuations.
However, the Poisson distribution approaches the Gaussian distribution in the large number limit.
Therefore, we can approximately use the Sheth-Tormen mass function for the oscillon number $\gtrsim 10$.
}
\begin{equation}
    \frac{\mathrm{d}n}{\mathrm{d}M}(z,M)
    =
    \frac{\bar{\rho}_{m}}{M} \frac{\mathrm{d}\ln\nu}{\mathrm{d}M} \nu f(\nu),
    \quad
    \left( \nu(z,M) \equiv \frac{\delta_c}{\sigma(z,M)} \right)
\end{equation} 
where $\bar{\rho}_{m}$ is the background matter density, $\delta_c = 1.686$ is the critical density contrast, and $f(\nu)$ is a probability density given by 
\begin{equation}
    \nu f(\nu)
    =
    2A \left( 1+\frac{1}{(a\nu)^{2p}} \right)
    \left( \frac{(a\nu)^{2}}{2} \right)^{1 / 2}
    \frac{\mathrm{e}^{-(a\nu)^{2} / 2}}{\sqrt{\pi}},
\end{equation}
with the constants $a=0.707, \, p=0.3,$ and $A = 0.322$.
$\sigma(z,M)$ is a smoothed mass variance over the scale $R(M)$ at redshift $z$, expressed as
\begin{equation}
    \sigma(z,M)^{2} 
    =
    \frac{1}{2 \pi^2}\int P(k,z)|W(k R)|^{2} \mathrm{d} k.
    \label{eq: variance_delta}
\end{equation}
Here the scale $R(M)$ is related to the minihalo mass $M$ by the relation $M=4 \pi R^3 \bar{\rho}_{m}/3$, and $W(kR)$ is the Fourier transform of the real space top-hat window function.

The mass functions for various cases are shown in Fig.~\ref{fig:massfunction}.
When we do not consider the oscillons, the number of minihalos is smaller than in the CDM case because the quantum pressure of ULAP suppresses the fluctuations on small scales.
On the other hand, when the oscillons are present, the matter power spectrum is enhanced, which increases the number of minihalos compared to the case without oscillons. 
We see that the mass function is even larger than that in the CDM case for heavier minihalos.

\begin{figure}[t]
    \begin{minipage}[b]{1\linewidth}
        \centering
        \includegraphics[clip,width=13cm]{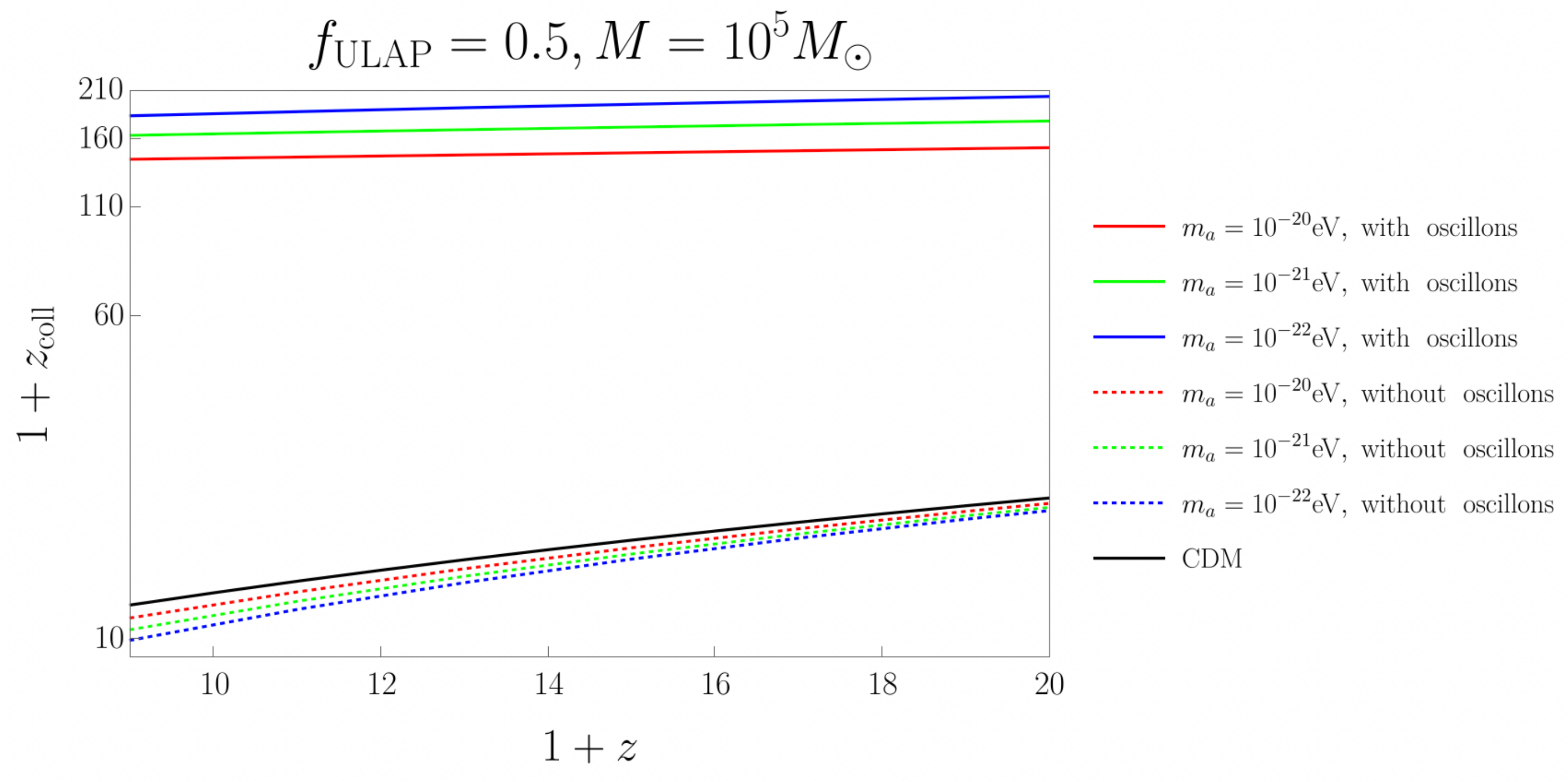}
     \end{minipage} \\
     \begin{minipage}[b]{1\linewidth}
        \centering
        \includegraphics[clip,width=13cm]{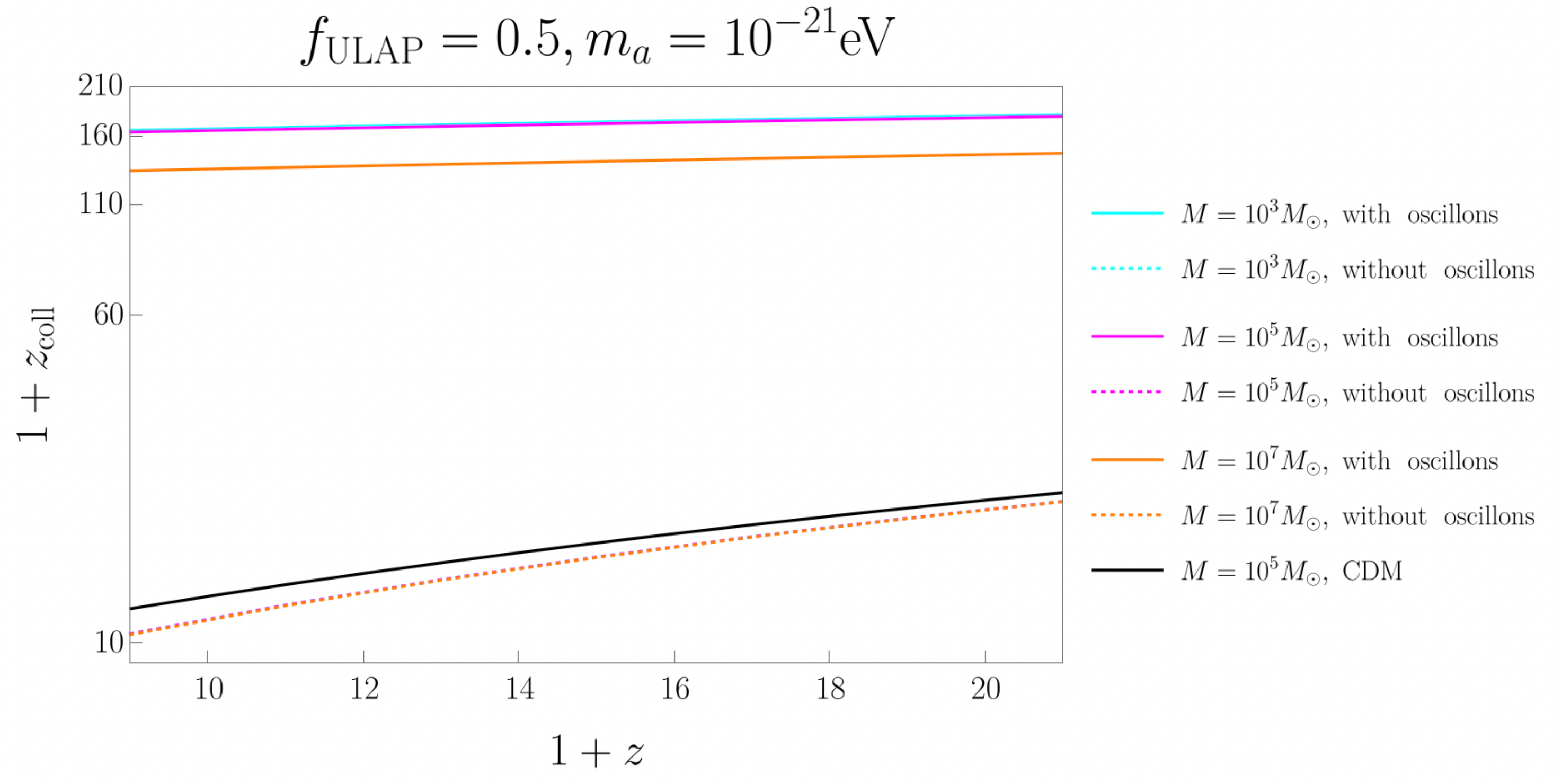}
     \end{minipage}
    \caption{The collapse redshift  $z_{\mathrm{coll}}$~\eqref{eq: zcoll} are shown as the function of redshift $z$ in the region of our interest for $f_{\mathrm{ULAP}}=0.5$.
    Line styles of the top panel is the same as in Figs.~\ref{power_spectrum} and \ref{fig:massfunction}.
    On the bottom panel, line styles represent the different masses of the minihalo.
    In the case of the model without oscillons, the values of the collapse redshift are too close together to distinguish.
    For the CDM case, only $M = 10^5 M_{\odot}$ was plotted since the change in mass did not change $z_{\mathrm{coll}}$ so much.
    }
    \label{fig:zcoll}
\end{figure}

Here, we consider the collapse time of each minihalo.
Since the density contrast is a random Gaussian variable
with variance $\sigma(z,M)$~\eqref{eq: variance_delta}
, the redshift at the halo collapse, $z_{\mathrm{coll}}$, also fluctuates.
According to the spherical collapse model, the halo collapses when the linear density contrast $\delta(z,M)$ exceeds the critical value $\delta_c$, so the collapse redshift is expressed as~\cite{Mo:1995cs}
\begin{equation}
    1+z_{\mathrm{coll}}
    =
    \frac{\delta(z)}{\delta_{c}}(1+z).
\end{equation}
Using this relation,
we can calculate the averaged collapse redshift as~\cite{Sekiguchi:2014wfa}
\begin{align}
    \left\langle 1+z_{\text {coll}} \right\rangle (M, z) 
    &=
    (1+z) 
    \frac{
        \int_{\delta_c}^{\infty} \mathrm{d} \delta \,
        \frac{\delta}{\delta_{c}} \frac{1}{ \sqrt{2 \pi} \sigma(z,M) }
        \exp \left[ -\frac{\delta^2}{2 \sigma(z,M)^2} \right]
    }{
        \int_{\delta_c}^{\infty} \mathrm{d} \delta \,
        \frac{1}{\sqrt{2 \pi} \sigma(z,M)} 
        \exp \left[ -\frac{\delta^2}{2 \sigma(z,M)^{2}} \right]
    } 
    \nonumber
    \\
    &=
    \left.
        (1+z) \frac{e^{-x^{2}}}{\sqrt{\pi} x \operatorname{erfc}(x)}
    \right|_{ x=\delta_{c} / \sqrt{2} \sigma(z,M) },
    \label{eq: zcoll}
\end{align}
where erfc($x$) represents the complementary error function. 
In the upper panel of Fig.~\ref{fig:zcoll}, we show $z_{\mathrm{coll}}$ as a function of $z$ for $f_{\mathrm{ULAP}}=0.5$ and a fixed minihalo mass $M=10^5M_{\odot}$.
Also, in the lower panel, we plot $z_{\mathrm{coll}}$ with various minihalo masses for a fixed ULAP mass $m_a=10^{-21}~\mathrm{eV}$.
Without oscillons, $z_{\mathrm{coll}}$ hardly depends on the minihalo mass.
In the $\Lambda$CDM model and ULAP model without oscillons, $\langle z_{\mathrm{coll}} \rangle \simeq z$ because $\sigma(z,M)$ is small compared to $\delta_c$.
On the other hand, in the case of the ULAP model with oscillons, $\sigma(z,M)$ is large compared to $\delta_c$,
and thus,  $\langle z_{\mathrm{coll}}\rangle$ is much larger than $z$.
In this case, the small halos were formed at a much earlier epoch as we can see in the lower panel of Fig.~\ref{fig:zcoll}.
In the following, physical quantities related to the halo will be calculated using averaged collapse redshift, which is denoted by $z_{\mathrm{coll}}$ for simplicity.

\subsection{Minihalo Profile}

In order to calculate the 21~cm fluctuations, we need to know the distribution of neutral hydrogen inside a minihalo, which absorbs and emits the 21~cm background photons. 
First, we review the DM profile, which is the source of the gas profile.
N-body simulations show that the DM distribution is described by the Navarro-Frenk-White (NFW) profile~\cite{Navarro:1996gj,Hennawi:2005bm},
\begin{equation}
    \rho_{\mathrm{DM}}(r)
    =
    \frac{\rho_{\mathrm{DM} 0}}{ \frac{r}{r_{s}}\left(1+\frac{r}{r_{s}}\right)^{2} },
\end{equation}
where $r_s$ is the scale radius, and $\rho_{\mathrm{\mathrm{DM} 0}}$ is the central density.
The scale radius is related to the virial radius $r_\mathrm{vir}$ through the concentration parameter $y \equiv r_{\mathrm{vir}}/r_s$, which is fitted by~\cite{Comerford:2007xb}
\begin{equation}
    y
    =
    \frac{14.8}{1+z_{\mathrm{coll}}}
    \left(\frac{M}{1.3 \times 10^{13} h^{-1} M_{\odot}}\right)^{-0.14},
\end{equation}
where $h=0.68$ is the Hubble parameter in units of $100$~km/sec/Mpc. 
The virial radius is calculated by the spherical collapse model~\cite{Barkana:2000fd} as 
\begin{equation}
    r_{\mathrm{vir}} 
    =
    0.784\, h^{-1}~\mathrm{kpc}
    \left(\frac{M}{10^{8} h^{-1} M_{\odot}}\right)^{1 / 3}
    \left(
        \frac{\Omega_{m 0}}{\Omega_{m}(z_{\mathrm{coll}})}
        \frac{\Delta}{18 \pi^{2}}
    \right)^{-1 / 3}
    \left( \frac{1+z_{\mathrm{coll}}}{10} \right)^{-1},
\end{equation}
where $\Delta \equiv 18 \pi^{2}+82 d-39 d^2$.
Here, $\rho_{m}(z)$ and $\Omega_m(z) \equiv \rho_{m}(z)/\rho_{c}(z)$ are the matter density and density parameter at redshift $z$, respectively.
Using these parameters, the central energy density is written as
\begin{equation}
    \rho_{\mathrm{DM} 0}
    =
    \frac{M}
    {
        \int_{0}^{r_{\mathrm{vir}}}
        \left[
            y \frac{r}{r_{\mathrm{vir}}}
            \left( 1+y \frac{r}{r_{\mathrm{vir}}} \right)^2
        \right]^{-1} \mathrm{d}^3r
    }.
\end{equation}

Next, we consider the halo gas profile, which is determined by the DM profile in the halo.
For simplicity, we make two assumptions: the gas within a halo is isothermal and in hydrostatic equilibrium.
Isothermal properties are guaranteed by the fact that the gas in the halo is in virial equilibrium.
Then, we can define the virial temperature $T_\mathrm{vir}$ as $\langle K \rangle = (3/2) k_B T_{\mathrm{vir}}$, where $\langle K \rangle$ is the time-averaged kinetic energy per particle and $k_B$ is the Boltzmann constant.
From this relation, the virial theorem leads to 
\begin{align}
    T_{\mathrm{vir}} 
    &=
    \frac{\mu m_{p}}{2 k_B}\frac{G M}{r_{\mathrm{vir}}}  
    \nonumber\\
    &=
    2.7 \times 10^4\, h^{-1}~\mathrm{K}
    \left(\frac{M}{10^{8} h^{-1} M_{\odot}}\right)^{1 / 3}
    \left(
        \frac{\Omega_{m 0}}{\Omega_{m}(z_{\mathrm{coll}})}
        \frac{\Delta}{18 \pi^{2}}
    \right)^{-1 / 3}
    \left( \frac{1+z_{\mathrm{coll}}}{10} \right)^{-1}, 
\end{align}
where $\mu=1.22$ is the mean molecular weight of the neutral hydrogen gas, and $m_p$ is the proton mass~\cite{Barkana:2000fd}. 
On the other hand, hydrostatic equilibrium means that the gas pressure $P(r)$ and gravitational force are balanced, that is
\begin{equation}
    \frac{\mathrm{d} P(r)}{\mathrm{d} r}
    =
    -\frac{G M_{\mathrm{in}}(r)}{r^{2}} \rho_{g}(r),
\end{equation}
where $\rho_g(r)$ is the gas density at a distance $r$ from the center and $M_{\mathrm{in}}(r)$ is the mass within a radius $r$~\cite{Makino:1997dv}.
Combined with the equation of state,
\begin{equation}
    P(r)=\frac{\rho_{g}(r)}{\mu m_p} k_{B} T_{\text {vir}},
\end{equation}
the gas density is obtained as 
\begin{equation}
    \rho_{g}(r)
    =
    \rho_{g 0} \exp \left[
        -\frac{\mu m_p}{2 k_{B} T_{\mathrm{vir}}}
        \left( v_{\mathrm{esc}}(0)^{2}-v_{\mathrm{esc}}(r)^{2} \right)
    \right],
\end{equation}
where $v_{\mathrm{esc}}(r)$ is the escape velocity of the halo given by
\begin{equation}
    v_{\mathrm{esc}}(r)^{2}
    =
    2 \int_{r}^{\infty} \frac{G M(\tilde{r})}{\tilde{r}^{2}} \mathrm{d} \tilde{r}
    =
    \frac{2 G M}{r} 
    \frac{\log (1+y \frac{r}{r_{\mathrm{vir}}} )}{ \left[ \log (1+y)-\frac{y}{1+y} \right] }.
\end{equation}
The central density $\rho_{g\mathrm{0}}$ is determined by requiring that the ratio of the baryon mass to the total matter mass in the halo is equal to $\Omega_b/\Omega_m$,
which leads to
\begin{equation}
    \rho_{g 0}
    =
    \frac{\Delta}{3} \frac{y^{3} e^{A}}
    { \int_{0}^{y}(1+t)^{A / t} t^{2} \mathrm{d} t }
    \frac{\Omega_{b}}{\Omega_{m}} \rho_{m}(z),
\end{equation}
where 
\begin{equation}
    A \equiv \frac{3 y}{\log (1+y)- \frac{y}{1+y}}.
\end{equation}
This gives us the number density of neutral hydrogen as
\begin{equation}
    n_\mathrm{H I}(r) = (1-Y) \frac{\rho_{g}(r)}{m_{p}},
\end{equation}
where $Y \simeq 0.25$ is the helium fraction. 

\begin{figure}[tb]
    \begin{minipage}[b]{0.5\linewidth}
        \centering
        \includegraphics[clip,scale=0.34]{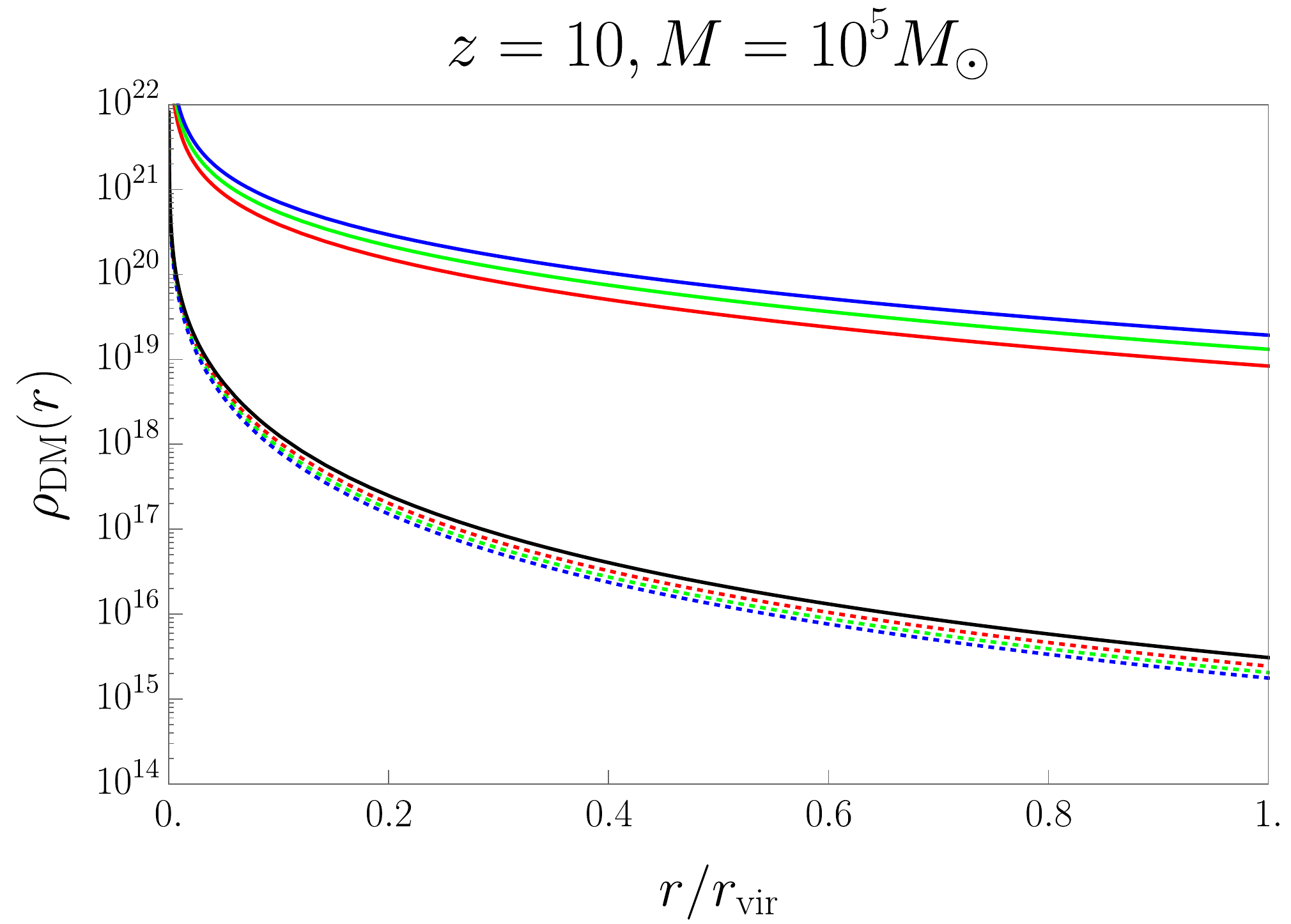}
    \end{minipage} 
    \begin{minipage}[b]{0.45\linewidth}
        \centering
        \includegraphics[clip,scale=0.34]{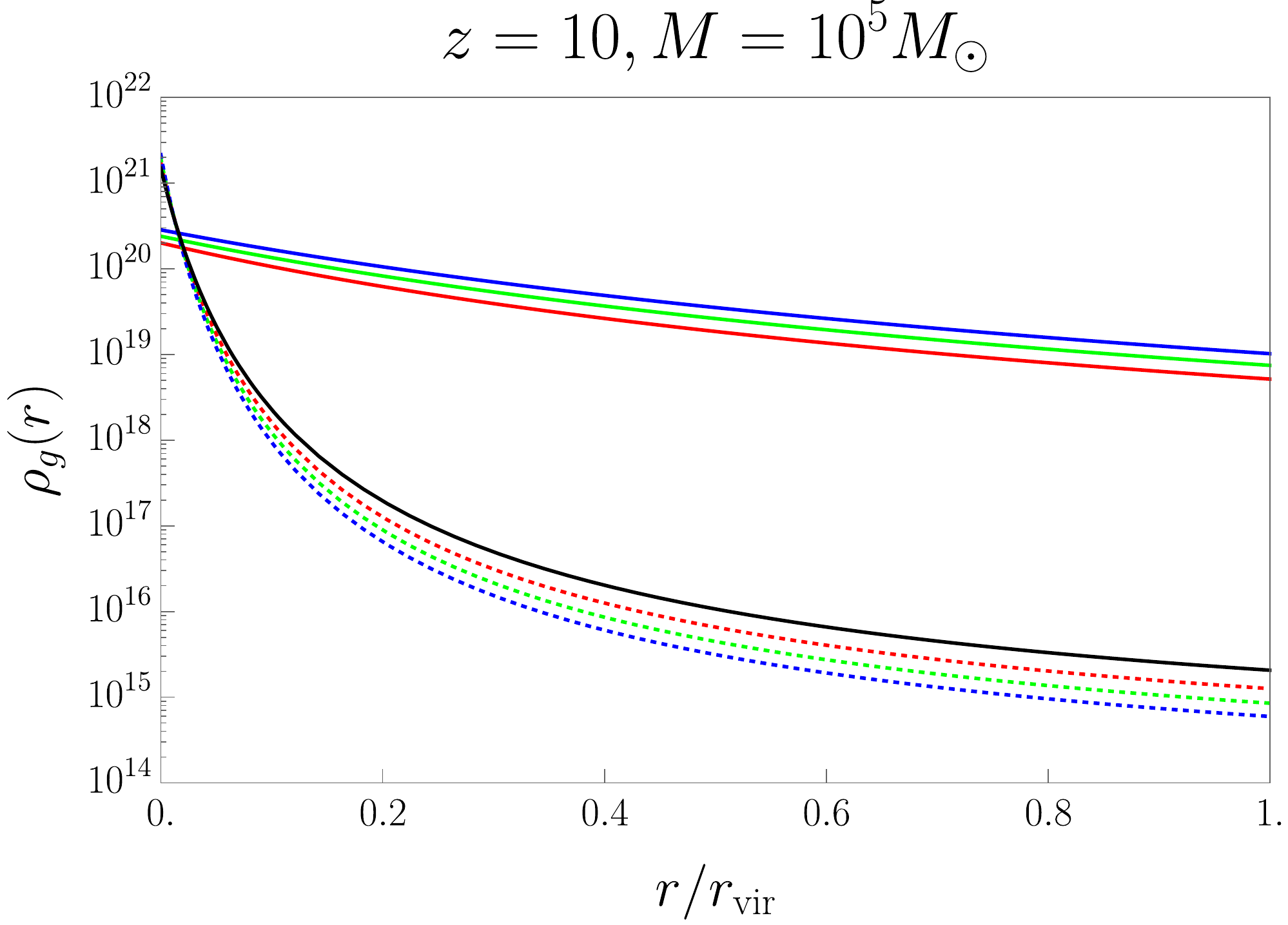}
    \end{minipage}
    \caption{Density profile of DM (left) and baryon gas (right) as a function of the normalized halo radius at $z=10$ for $M=10^5M_{\odot}$.
     Line styles are the same as in Figs.~\ref{power_spectrum} and \ref{fig:massfunction}.}
\label{profile}
\end{figure}

The DM and gas profiles are shown in Fig.~\ref{profile}.
It is seen that the gas profile does not follow the DM profile in particular near the center.
This contrasts with the truncated isothermal sphere (TIS) model, which is commonly used in other papers~\cite{Shapiro:1998zp,Iliev:2001he}.
We will discuss this point later.

\subsection{Spin Temperature}

The 21~cm line photons are emitted and absorbed by the transition between the ground state and the excited state of the hyperfine structure of hydrogen atom.
The spin temperature is defined using the number density ratio of the excited state to the ground state~\cite{4065250}:
\begin{equation}
    \frac{n_{1}}{n_{0}} 
    \equiv 
    \frac{g_{1}}{g_{0}} \exp \left(-\frac{T_\ast}{T_{s}}\right),
\end{equation}
where 
$n_1$ and $n_0$ are the number density of hydrogen atoms in the excited and ground states, $g_1=3$ and $g_0=1$ are the degrees of freedom of the excited and ground states respectively, 
and $T_\ast \simeq 68 ~\mathrm{mK}$ is the temperature corresponding to the energy difference between the two states.

\begin{figure}[tb]
    \centering
    \includegraphics[clip,width=15cm]{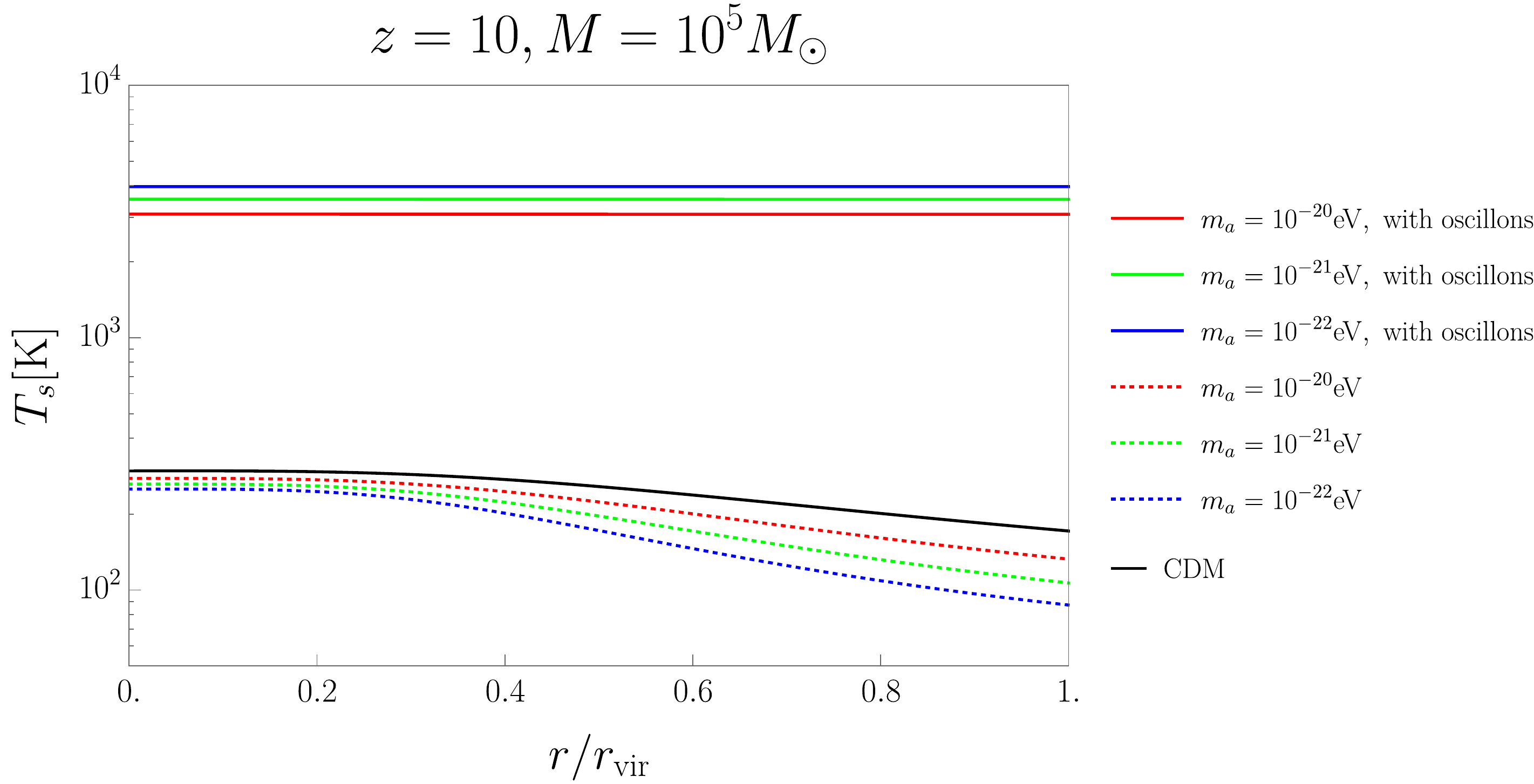}
    \caption{Spin temperature in a minihalo with $M=10^5M_{\odot}$ as a function of the normalized halo radius at $z=10$.
    The difference in line styles is the same as in Figs.~\ref{power_spectrum} and \ref{fig:massfunction}.}
    \label{spin_temperature}
\end{figure}

The physical processes involved in the transition between the two states are spontaneous emission, absorption, and stimulated emission.
The transition rate for spontaneous emission is given by $A_{10}=2.85 \times 10^{-15}~\mathrm{s}^{-1}$.
The transition rates of the other two processes are determined by the details of three interactions; absorption and  emission of a CMB photon, particle collision, and the Wouthuysen-Field effect triggered by a Ly-$\alpha$ photon.
Assuming that all these three processes are in equilibrium, the spin temperature can be written as a weighted average of the three temperatures corresponding to the three processes,
\begin{equation}
\label{eq:spin}
    T_{s}
    = 
    \frac{T_{\gamma}+y_{c} T_{K}+y_{\alpha} T_{\alpha}}{1+y_{c}+y_{\alpha}}, 
\end{equation}
where $T_\gamma,\,T_K,$ and $T_\alpha$ are the CMB photon temperature, kinetic (or gas) temperature, and Ly-$\alpha$ color temperature respectively, and $y_c$, $y_\alpha$ are the normalized deexcitation rate of particle collision and Ly-$\alpha$ pumping, respectively~\cite{Madau:1996cs}.
Here, we use the virial temperature $T_{\mathrm{vir}}$ as the gas temperature $T_K$, which is justified when cooling is not efficient.
Moreover, since there are few free electrons at the time we are interested in, we consider only H-H collisions for particle collisions.
We can, then, write the normalized deexcitation rate of particle collision $y_c$ as
\begin{equation}
    y_c
    = 
    \frac{T_\ast n_{\mathrm{HI}} \kappa_{10}}{A_{10} T_{\mathrm{K}}},
\end{equation}
where $\kappa_{10}$ is the deexcitation rate of H-H collision, approximated by~\cite{Kuhlen:2005cm} 
\begin{equation}
    \kappa_{10}
    =
    3.1 \times 10^{-11}
    \left( \frac{T_{K}}{1 \mathrm{~K}} \right)^{0.357}
    \exp \left(-\frac{32 \mathrm{~K}}{T_{K}} \right) 
    ~\mathrm{cm}^{3} \mathrm{~s}^{-1}.
\end{equation}
As for $y_\alpha$, we set $y_\alpha = 0$ because, at the time we are considering, the process of star formation depends on astrophysics and contains large uncertainties.
Thus, we do not consider the contribution from Ly-$\alpha$, which primarily comes from stars.\footnote{
Since it is often the case that $T_\alpha \approx T_K$~\cite{Furlanetto:2006jb}, the weight of $T_K$ in the Eq.~\eqref{eq:spin} becomes larger when considering the Wouthuysen-Field effect.
This will make the spin temperature $T_s$ close to kinetic temperature $T_K$.
}
The spin temperatures $T_s$ for various models are shown in Fig.~\ref{spin_temperature}.
It is seen from Fig.~\ref{spin_temperature} that the spin temperature is nearly independent of $r$ when oscillons are present.
This is because large $z_{\mathrm{coll}}$ makes $y_c$ larger, which leads to $T_s \simeq T_{\mathrm{K}}$.
In the absence of oscillons, the spin temperature is small compared to the CDM model due to the suppression of fluctuation
by ULAP, which leads to smaller $z_{\mathrm{coll}}$.

\subsection{Brightness Temperature}

Now let us consider the effects of minihalos on the 21~cm radiation.
The observable related to the 21~cm radiation is the brightness temperature, which is the temperature of the incoming flux when the radiation is assumed to be blackbody radiation in the Rayleigh-Jeans limit. The brightness temperature of the 21~cm line passing through a minihalo of mass $M$ with an impact parameter $\alpha$ is estimated by solving the radiative transfer equation, which leads to
\begin{equation}
    T_{b}\left(\nu, M, \alpha \right)
    =
    T_{\gamma}(z) e^{-\tau_{\mathrm{max}}(\nu, M, \alpha)} 
    + \int_{-\infty}^\infty \mathrm{d}x \,
    T_{\mathrm{s}}(r) e^{-\tau(\nu, M, \alpha, x)} 
    \frac{\partial \tau(\nu, M, \alpha, x)}{\partial x},
\end{equation}
where 
$|x|=\sqrt{r^2-\alpha^2}$ is the length along the line of sight.
The partial optical depth $\tau(\nu, M, \alpha, x)$ and the total optical depth $\tau_{\mathrm{max}}(\nu, M, \alpha)$ through a minihalo along the line of sight are calculated by integrating the absorption coefficient as~\cite{2004}
\begin{align}
    \tau(\nu, M, \alpha, x)
    &=
    \frac{3 c^{2} A_{10} T_\ast}{32 \pi \nu_{\ast}^2}
    \int_{-x_{\mathrm{max}}}^{x} \frac{n_\mathrm{H I}(r)}{T_{s}(r)} \phi(\nu) \mathrm{d} x 
    ,
    \\
    \tau_{\mathrm{max}}(\nu, M,\alpha)
    &=
    \tau \left(\nu, M, \alpha, x_{\mathrm{max}} \right)
    ,
\end{align}
where $\nu_\ast = 1420~\mathrm{MHz}$ is the frequency of the 21~cm emission, $x_{\mathrm{max}}=\sqrt{r_{\mathrm{vir}}^2-\alpha^2}$, and $\phi(\nu)$ is the Doppler broadening line profile written as  
\begin{equation}
    \phi(\nu)
    =
    \frac{1}{\sqrt{\pi} \Delta \nu} 
    \exp \left[ -\frac{\left(\nu -\nu_\ast\right)^2}{\Delta \nu^{2}} \right]
    ,
    \quad
    \left( 
        \Delta \nu^2 \equiv \nu_\ast \frac{2 k_{B} T_{\mathrm{vir}}}{m_{p}c^2}
    \right)
    .
\end{equation}

Since what we can observe is the difference between the brightness temperature and CMB temperature, we define the differential brightness temperature with respect to the CMB as
\begin{equation}
    \Delta T_b \equiv \frac{\langle T_b \rangle - T_\gamma(z)}{1 + z},
\end{equation}
where $\langle T_b \rangle\equiv \int T_b(\nu_\ast, M, \alpha) \, \mathrm{d}A / A$ is the averaged brightness temperature over the halo cross section $A \equiv \pi r_{\mathrm{vir}}^2$.
Furthermore, we take a weighted average of $\Delta T_b$ over the mass distribution.
The mean differential brightness temperature is then written as~\cite{2002} 
\begin{equation}
    \overline{\Delta T}_{b}
    =
    \frac{c(1+z)^{4}}{\nu_{\ast} H(z)} 
    \int_{M_{\min }}^{M_{\max }} \Delta \nu_{\mathrm{eff}} \Delta T_{b}A 
    \frac{\mathrm{d} n}{ \mathrm{d} M} \mathrm{d} M,
\end{equation}
where $\Delta \nu_{\mathrm{eff}} = [\phi(\nu_{\ast})(1+z)]^{-1}$ is the effective redshifted linewidth, and $M_{\mathrm{max}}$ and $M_{\mathrm{min}}$ are maximum and minimum minihalo masses, respectively.
We set $M_{\mathrm{min}}$ as the Jeans mass~\cite{Barkana:2000fd}
\begin{equation}
    M_{\mathrm{min}}(z)
    =
    5.73 \times 10^{3}\left(\frac{\Omega_{m} h^{2}}{0.15}\right)^{-1/2}
    \left(\frac{\Omega_{b} h^{2}}{0.02}\right)^{-3 / 5}
    \left(\frac{1+z}{10}\right)^{3 / 2} M_{\odot},
\end{equation}
and $M_{\mathrm{max}}$ is determined by the condition $T_{\mathrm{vir}}(M)<10^4\, \mathrm{K}$, for which the star formation is not effective due to the inefficient atomic gas cooling. 

Since the spatial fluctuations of the minihalo distribution are proportional to the density fluctuations~\cite{Mo:1995cs}, the fluctuations of the brightness temperature should be written as a product of the density fluctuations and the brightness temperature with a bias factor.
From the above relations, we can derive the root-mean-square (rms) of the 21~cm signal corresponding to $3 \sigma$ fluctuations as
\begin{equation}
\label{eq:rms}
    \left\langle\delta T_{b}^{2}\right\rangle^{1 / 2}
    =
    3 \sigma_{p}\left(\Delta \theta, \Delta \nu_{\mathrm{b}}\right)
    \bar{b}(z) \overline{\Delta T_{b}},
\end{equation}
where $\sigma_p(\Delta \theta , \Delta \nu_{\mathrm{b}})$ is the mass variance filtered over cylindrical pencil beam with the angular size $\Delta \theta$ and frequency band width $\Delta\nu_{\mathrm{b}}$.
Using comoving radius $R=\Delta \theta(1+z) D_{A}(z) / 2$ and length $L \approx(1+z) c H(z)^{-1}(\Delta \nu_{\mathrm{b}} / \nu)$ of the beam,
$\sigma_p$ is expressed as~\cite{Tozzi:1999zh,2002}
\begin{equation}
    \sigma_{p}^{2}
    =
    \frac{8 D^{2}(z)}{\pi^{2} R^{2} L^{2}} 
    \int_{0}^{\infty} \mathrm{d} k \int_{0}^{1} \mathrm{d} x \,
    \frac{
        \sin ^{2}\left( \frac{k L x}{2} \right) 
        J_{1}^{2}\left[
            k R\left(1-x^{2}\right)^{1 / 2} 
        \right]
    }
    {x^{2}\left(1-x^{2}\right)}
    \left(1+f x^{2}\right)^{2} \frac{P(k)}{k^{2}},
\end{equation}
where $D(z)$ is a linear growth factor normalized by the condition $D(0)=1$,
$J_1$ is the Bessel function,
and $f \approx \Omega_{\mathrm{m}}^{0.6}$~\cite{1987MNRAS:2271K}.
The factor $(1+f x^2)^2$ is included because the peculiar velocities change the effects of Hubble expansion slightly, which leads to the correction to the cylinder length.
$\bar{b}(z)$ is the flux-weighted bias factor defined by
\begin{equation}
    \bar{b}(z)
    \equiv
    \frac{
        \int_{M_{\min }}^{M_{\max }} \mathrm{d} M \,
        \frac{\mathrm{d} n}{\mathrm{d} M} \mathcal{F}(z, M) b(z,M)
    }{
        \int_{M_{\min }}^{M_{\max }} \mathrm{d} M \,
        \frac{\mathrm{d} n}{\mathrm{d} M} \mathcal{F}(z, M)
    },
\end{equation}
%
where $\mathcal{F}(z,M)=\langle T_{\mathrm{b}} \rangle A \sigma_v$ is the effective flux from a minihalo, 
$\sigma_v$ is the velocity dispersion of a minihalo, 
and 
\begin{equation}
    b\left(z,M\right) 
    =
    1+\frac{\nu(z,M)^2-1}{\delta (z)}
\end{equation}
is a halo bias, which represents how the minihalo distribution is biased relative to the matter distribution~\cite{Mo:1995cs}.

\section{21~cm signals and sensitivity}
\label{sec4}

\begin{figure}[t]
    \begin{tabular}{cc}
      \begin{minipage}[t]{0.45\hsize}
        \centering
        \includegraphics[clip,scale=0.33]{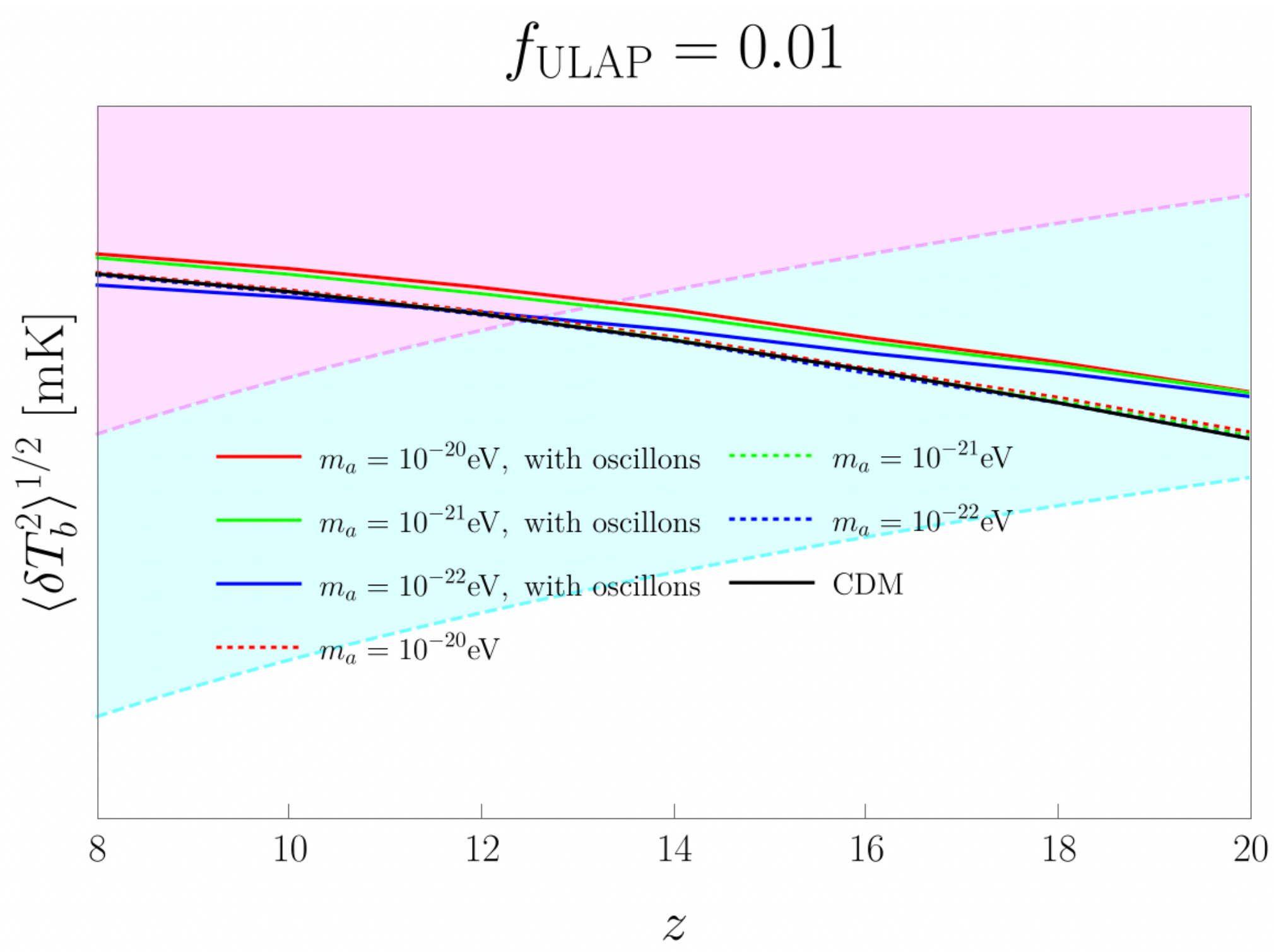}
      \end{minipage} &
      \begin{minipage}[t]{0.45\hsize}
        \centering
        \includegraphics[clip,scale=0.33]{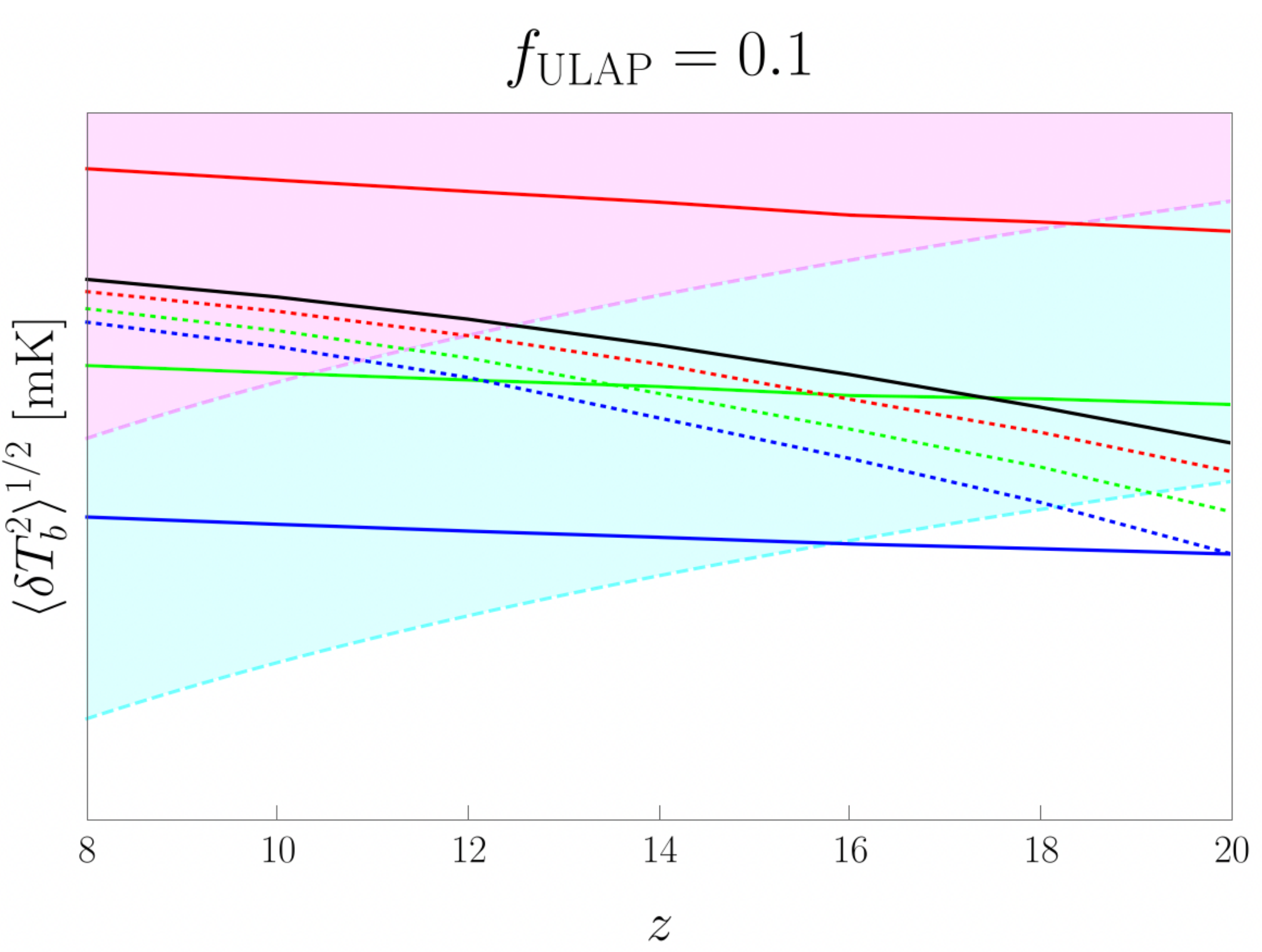}
      \end{minipage} \\
   
      \begin{minipage}[t]{0.45\hsize}
       \centering
        \includegraphics[clip,scale=0.33]{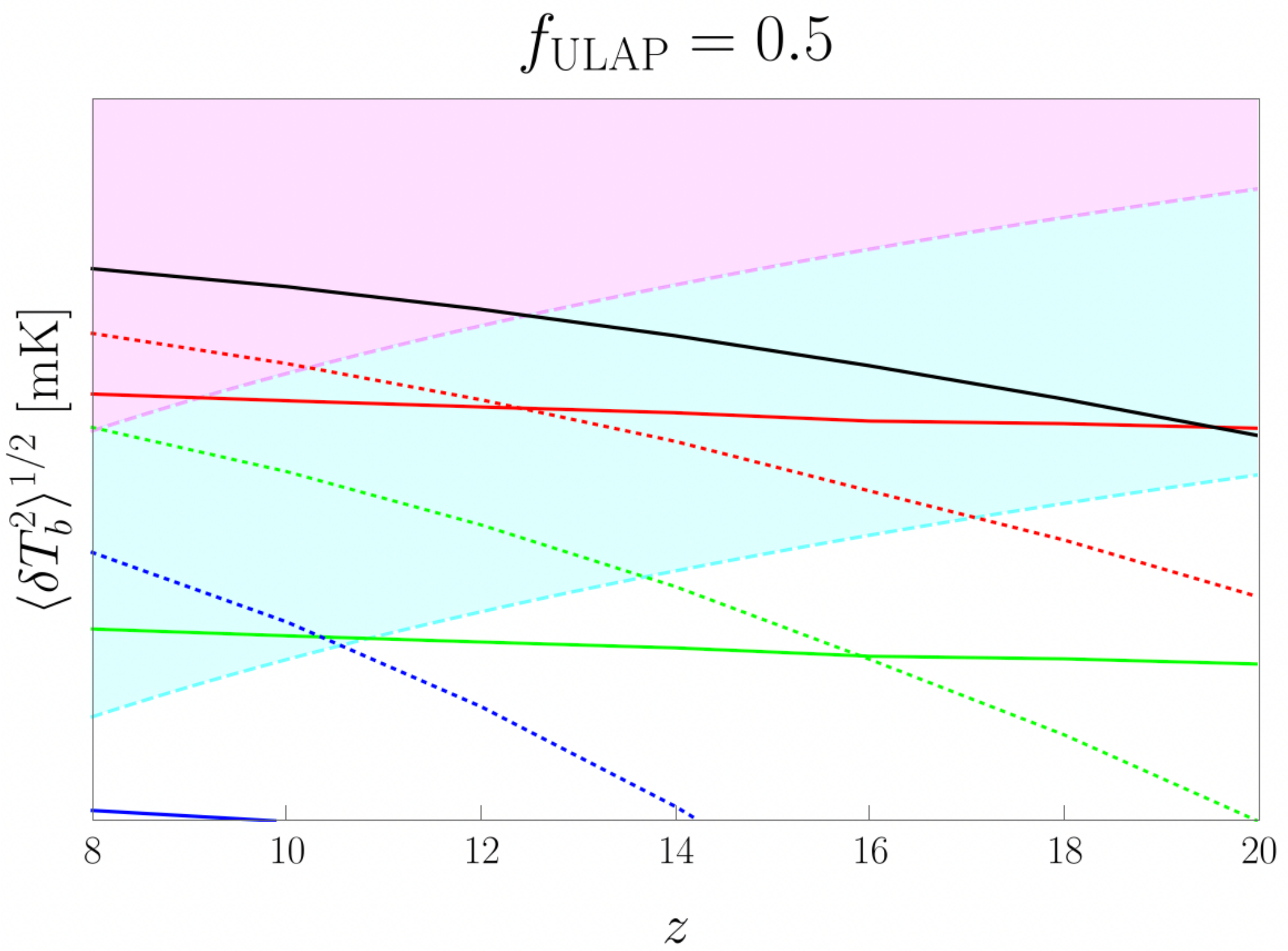}
      \end{minipage} &
      \begin{minipage}[t]{0.45\hsize}
        \centering
        \includegraphics[clip,scale=0.33]{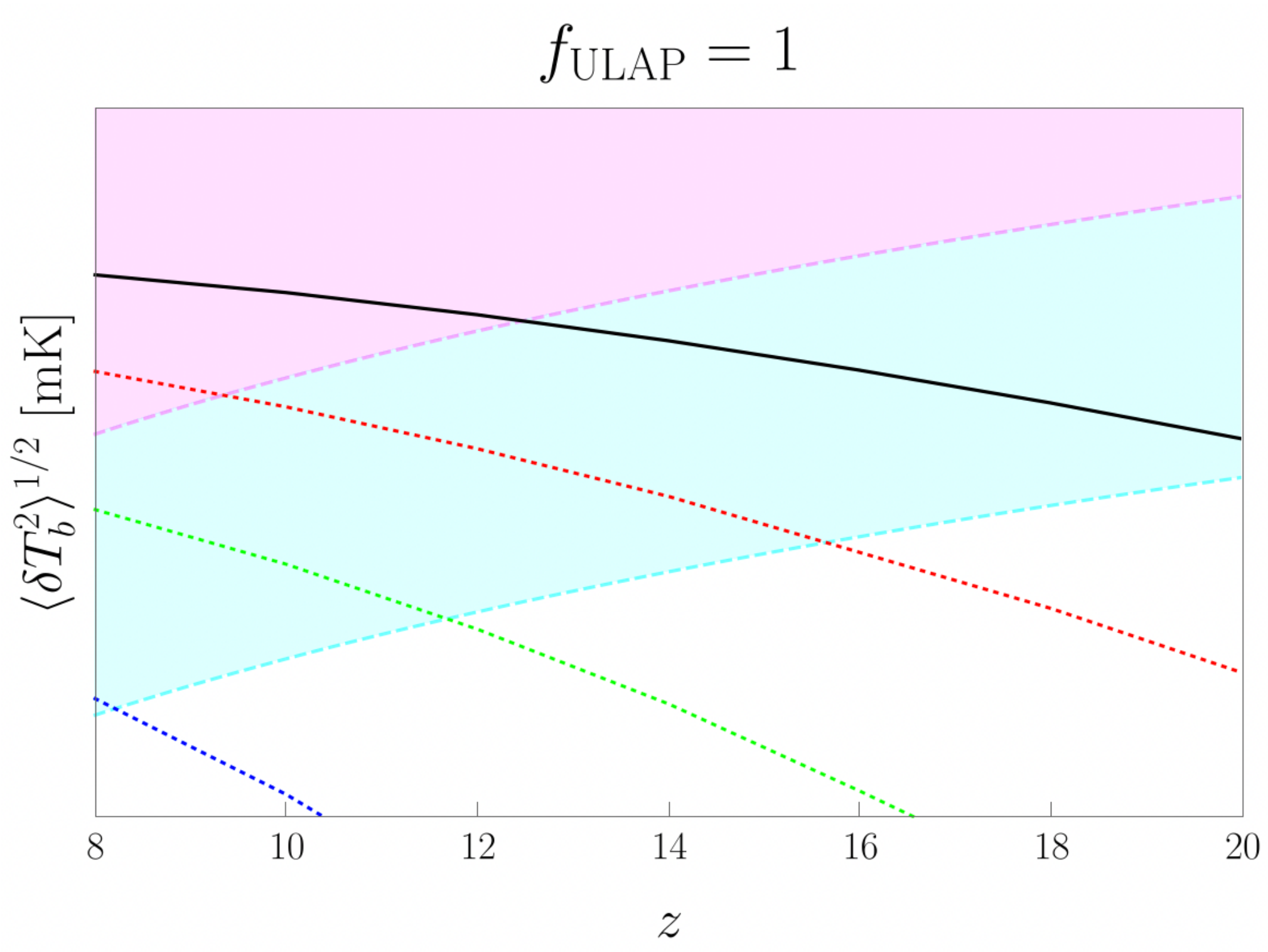}
      \end{minipage} 
    \end{tabular}
     \caption{Root-mean-square of the 21~cm fluctuation as a function of redshift.
     We have taken $f_{\mathrm{ULAP}}=0.01, \, 0.1, \, 0.5, \, 1$ from left top to bottom right.
     The regions shaded in light blue and magenta show the sensitivity curve of SKA and FFTT-like observation.
     Line styles are the same as in Figs.~\ref{power_spectrum} and \ref{fig:massfunction}.}
     \label{result}
  \end{figure}

We evaluate the rms of the 21~cm fluctuation using Eq.~\eqref{eq:rms} as shown in Fig.~\ref{result}.
The red hatched and blue hatched regions represent the sensitivity of the SKA and Fast Fourier Transform Telescope (FFTT)-like observation, respectively.
The noise of these observations is given by~\cite{Furlanetto:2006jb,Tegmark:2008au}
\begin{equation}
    \delta T_{\text{noise}}
    =
    20~\mathrm{mK} \frac{10^{4} \mathrm{~m}^{2}}{A_{\mathrm{tot}}}
    \left(
        \frac{10~\operatorname{arcmin}}{\Delta \theta}
    \right)^{2}
    \left( \frac{1+z}{10} \right)^{4.6}
    \left( 
        \frac{\mathrm{MHz}}{\Delta \nu_{\text {b}}}
        \frac{100 h}{t_{\mathrm{int}}}
    \right)^{1 / 2},
\end{equation}
where $A_{\mathrm{tot}}$ is the effective collecting area, and $t_{\mathrm{int}}$ is the integration time.
Here, we set $A_{\mathrm{tot}}=10^5~\mathrm{m}^2$ for the SKA-like observation, and $A_{\mathrm{tot}}=10^7~\mathrm{m}^2$ for the FFTT-like observation.
The other parameters are set as $\Delta \theta=9~\mathrm{arcmin}$, $\Delta \nu_{\text {b}}=1~\mathrm{MHz}$, and $t_{\mathrm{int}}=1000 h$. 

From Fig.~\ref{result}, one can see that, in the case of ULAP without oscillons, the 21~cm fluctuation signal becomes weaker as $f_{\mathrm{ULAP}}$ increases for a given redshift.
This is because ULAP suppresses the matter fluctuations on small scales and reduces the number of minihalos that contribute to the 21~cm signal.
As the ULAP fraction decreases, the suppression becomes less effective and the signal asymptotically approaches that in the CDM case.
On the other hand, in the presence of oscillons, the 21~cm signal changes non-trivially as $f_{\mathrm{ULAP}}$ increases.
There is no signal for $f_{\mathrm{ULAP}}=1$ because the fluctuations are amplified so much that there is no minihalo whose virial temperature is smaller than $10^4~\mathrm{K}$, and the 21~cm fluctuations are not generated.
This is because we use $z_{\mathrm{coll}}$ for $z$ in the formulas of $M_{\mathrm{min}}$ and $M_{\mathrm{max}}$, leading to $M_{\mathrm{min}}> M_{\mathrm{max}}$, i.e. no mass range for halos inducing the 21~cm signal.
We will discuss uncertainties of this procedure in Sec.~\ref{sec5}.
As $f_{\mathrm{ULAP}}$ becomes smaller, the signals first become larger and exceed the CDM case and then approach the CDM case, which is due to the competing effects of the increasing fraction of minihalos contributing to the 21~cm fluctuations and 
the decreasing total number of minihalos
as $f_{\mathrm{ULAP}}$ decreases.


\begin{figure}[t]
    \centering
    \includegraphics[clip,width=15cm]{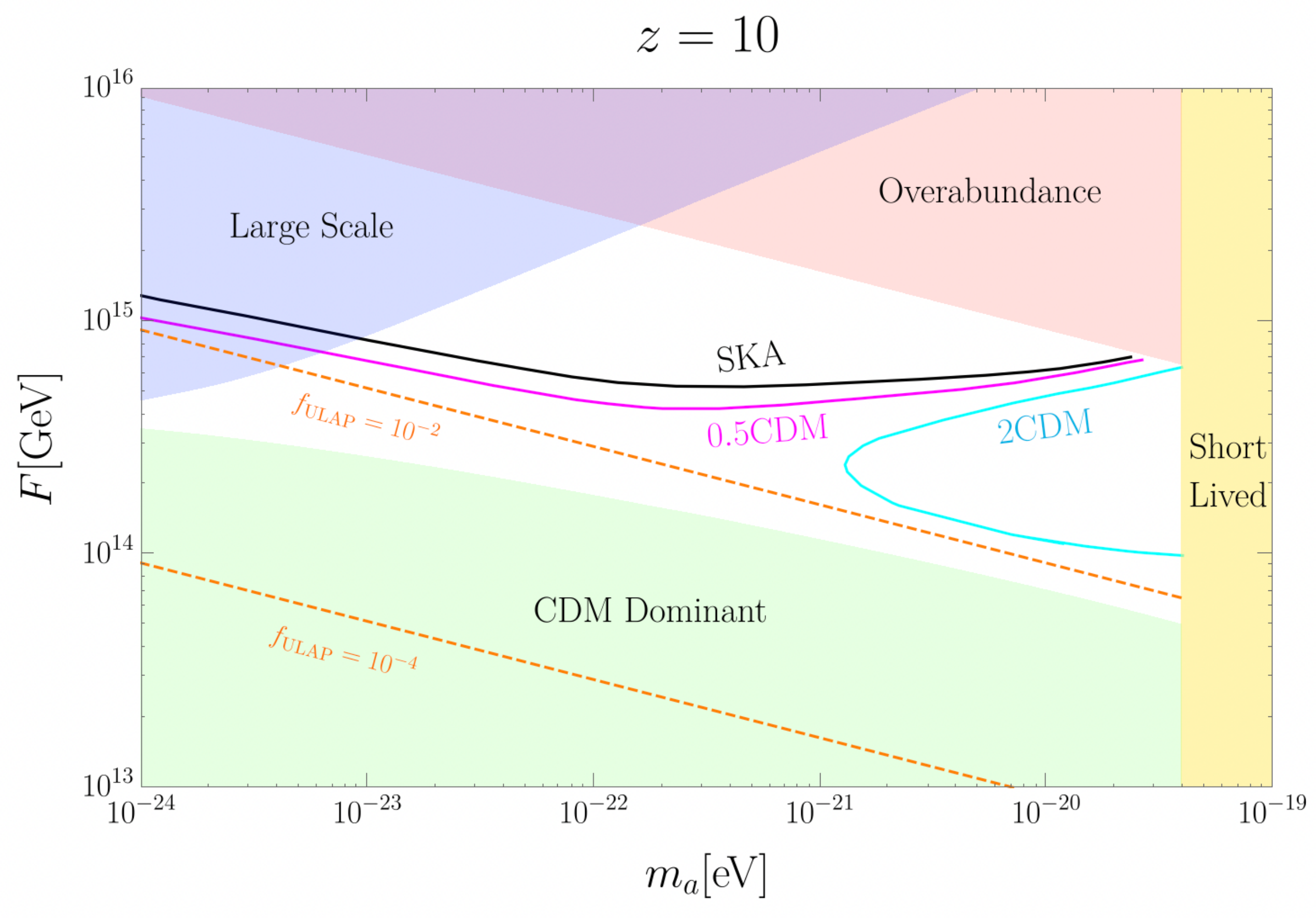}
    \caption{Parameter region where the 21 cm signal is observed in future observation by SKA. 
    The vertical and horizontal axes represent the ULAP decay constant and ULAP mass, respectively. 
    The black, magenta, and cyan lines correspond to the sensitivity of SKA, half of the CDM signal, and twice as large as the CDM signal, respectively.
    The region between the black and magenta lines and the region surrounded by the cyan line can be distinguished from the CDM signal.
    The blue shaded region is excluded by large-scale structure observations.
    The red shaded region indicates the overabundance of ULAP, $f_{\mathrm{ULAP}} > 1$.
    Within the green shaded region, it is not possible to obtain a signal distinguishable from the case without oscillons because the power spectrum of oscillons is smaller than the CDM and ULAP power spectrum at cut-off scale $k_c$, $P_{\mathrm{osc}}(k_c) < P_{\mathrm{CDM}}(k_c) + P_{\mathrm{ULAP}}(k_c)$.
    In the yellow region of $m_a > 4 \times 10^{-20}~\mathrm{eV}$, the oscillons decay before $z = 10$, so their effects cannot be observed in the 21~cm lines.}
    \label{fig:sensitivity}
\end{figure}

Fig.~\ref{fig:sensitivity} shows the parameter regions observable in future observations by SKA at redshift $z=10$. 
We chose $z = 10$ as the reference redshift because the signal is large enough to be observed and reionization does not occur.
The black solid line shows the upper bound of the ULAP decay constant $F$ below which the signal of 21 cm fluctuations is larger than the SKA sensitivity. 
The magenta and cyan lines indicate the parameters with which the signal with oscillons is twice and half of the CDM one, respectively. 
We roughly set the criterion that we can distinguish the signal with oscillons from the CDM one if it is larger than twice the CDM signal or smaller than half of the CDM signal.
Thus, the ULAP in the parameter region between the black line and the magenta line, or surrounded by the cyan line, is considered to be distinguished from CDM by the SKA observations in a rough estimation.

In addition to the observability of the signal of the 21~cm fluctuations, we consider the following constraints. 
The blue shaded region is excluded because the oscillon power spectrum is larger than $\Lambda$CDM one on large scale $k < \mathcal{O}(1)~\mathrm{Mpc}^{-1}$, $P_{\mathrm{osc}}(1/\mathrm{Mpc}) > P_{\Lambda \mathrm{CDM}}(1/\mathrm{Mpc})$~\cite{Planck:2018vyg,2017AJ....154...28B,2018ApJS..235...42A,DES:2017myr}.
The red shaded region is constrained by the condition $f_{\mathrm{ULAP}} \leq 1$.
We should note that this limit depends on the initial value of the ULAP.
The initial value of the scalar field is set as $\phi(t_{\mathrm{ini}})/F = 12 \pi$, but the limit becomes stricter if it becomes larger.
We cannot obtain a signal distinguishable from the signal in the case without oscillons in the green shaded region because the power spectrum of oscillons at $k_c$ is smaller than that of CDM and ULAP, $P_{\mathrm{osc}}(k_c) < P_{\mathrm{CDM}}(k_c) + P_{\mathrm{ULAP}}(k_c)$ (see Sec.~\ref{sec:oscillon_power}). 
In the yellow shaded region, which corresponds to the region of $m_a > 4 \times 10^{-20}$, the oscillons decay before $z=10$, so 
we cannot apply our analysis.

\section{Conclusion}
\label{sec5}

In this paper, we evaluated the rms of the 21~cm fluctuations due to minihalos in the presence of ULAP oscillons.
Since oscillons amplify the fluctuations on small scales, it has a non-trivial effect on the anisotropies of the 21~cm background; if the ULAP with oscillons dominates DM $f_{\mathrm{ULAP}} \sim 1$,
the number of minihalos that contribute to the 21~cm fluctuations decreases and the 21~cm fluctuations cannot be observed. 
On the other hand, for $f_{\mathrm{ULAP}} \sim 0.1$, the signal is larger than that in the CDM case, which could provide an evidence for the presence of oscillons if observed. 

We should make some comments on the uncertainties of our analysis.
We have applied the NFW profile to the DM distribution in a halo, while others use the TIS model, which assumes hydrostatic equilibrium between DM and baryons.
This assumption seems to be invalid, as can be read from Fig.~\ref{profile}.
On the other hand, the NFW profile may not describe the internal structure of a minihalo in the presence of oscillons because the mass of the produced oscillons can be comparable to the mass of the minihalo in our parameter region of interest.
This fact leads to more concentration of gas in the center of the  minihalo and results in larger 21~cm fluctuation signals. 

Furthermore, we used the averaged collapse redshift $\langle z_{\mathrm{coll}} \rangle$ in the calculation of the 21~cm fluctuation 
in evaluating the physical properties of minihalos.
However, it is more accurate to consider the collapse redshift as a random variable which obeys the Gaussian distribution.
This leads to an underestimation of the 21~cm fluctuation signal when $f_{\mathrm{ULAP}} \sim 1$ and oscillons are present.
With this taken into account, the 21~cm fluctuation signal may be non-zero even when $f_{\mathrm{ULAP}} \sim 1$ in the presence of oscillons.
Finally, we only consider the case where the oscillons survive until $z \sim 10$, but if they decay before then, the emitted ULAPs from oscillons, which behave like warm DM, can suppress the small-scale fluctuations and provide another constraint~\cite{Imagawa:2021sxt}.

\section*{Acknowledgements}
We would like to thank Shin Kobayashi for helpful advice
and Wakutaka Nakano for kindly permitting us to use the numerical table of the decay rate of oscillons.

This work is supported by the Grant-in-Aid for Scientific Research Fund of the JSPS 20H05851(M.\,K.), 21K03567(M.\,K.),
20J20248 (K.\,M.) and
19J21974 (H.\,N.).
M.\,K. and K.\,M. are supported by World Premier International Research Center Initiative (WPI Initiative), MEXT, Japan (M.\,K. and K.\,M.). 
K.\,M. is supported by the Program of Excellence in Photon Science.
H.\,N. is supported by Advanced Leading Graduate Course for Photon Science.
\bibliographystyle{JHEP}
\bibliography{bibtex}
\end{document}